\documentclass[preprint2]{aastex}
\usepackage{psfig}

\newcommand\mzon   {M$_{\odot}$}
\newcommand\pp     {$\pm$}

\newcommand\Lunit {erg s$^{-1}$}
\newcommand\funit {erg s$^{-1}$ cm$^{-2}$}
\newcommand\nh    {$N_{\rm H}$}

\righthead{A {\itshape XMM-Newton} observation of SAX J1808.4--3658}
\slugcomment{}

\begin{document}

\title{A {\itshape XMM-Newton} observation during the 2000 outburst of
SAX J1808.4--3658}

\author{Rudy Wijnands\footnote{Chandra Fellow} \footnote{Present address: 
School of Physics and Astronomy, University of St Andrews, North
Haugh, St Andrews, Fife, KY16 9SS, Scotland, UK;
radw@st-andrews.ac.uk}}

\affil{Center for Space Research, Massachusetts Institute of
Technology, 77 Massachusetts Avenue, Cambridge, MA 02139-4307, USA;
rudy@space.mit.edu}

\begin{abstract}
I present a {\it XMM-Newton} observation of the accretion driven
millisecond X-ray pulsar SAX J1808.4--3658 during its 2000
outburst. The source was conclusively detected, albeit at a level of
only $\sim2\times10^{32}$ \Lunit. The source spectrum could be fitted
with a power-law model (with a photon index of $\sim$2.2), a neutron
star atmosphere model (with a temperature of $\sim$0.2 keV), or with a
combination of a thermal (either a black-body or an atmosphere model)
and a power-law component.  During a {\it XMM-Newton} observation
taken approximately one year later, the source was in quiescence and
its luminosity was a factor of $\sim$4 lower. It is possible that the
source spectrum during the 2000 outburst was softer than its quiescent
2001 spectrum, however, the statistics of the data do not allow to
make a firm conclusion. The results obtained are discussed in the
context of the 2000 outburst of SAX J1808.4--3658 and the quiescent
properties of the source.

\end{abstract}


\section{Introduction\label{section:introduction}}

The X-ray transient SAX J1808.4--3658 was discovered in September 1996
when it exhibited a weak outburst lasting only a few weeks
\citep{intzandetal1998,intzandetal2001}.  In April 1998 the source was
found to be in outburst again \citep{marshall1998} and it was
discovered that the source exhibits coherent millisecond X-ray
oscillations with a frequency of approximately 401 Hz
\citep{wvdk1998}. In early 2000, the source exhibited a third outburst
during which it showed erratic luminosity behavior with luminosity
swings of three orders of magnitude within a few days
\citep{wijnandsetal2001_rxte,wijnandsetal2002_bepposax}. This erratic
behavior lasted for several months before the source returned to
quiescence. Very recently, in October 2002, a fourth outburst of the
source was detected \citep{markwardtetal2002} during which its peak
luminosity was very similar to that observed during the 1996 and 1998
outbursts.

In quiescence, SAX J1808.4--3658 has been observed on several
occasions with the {\it BeppoSAX} and {\it ASCA} satellites
\citep{stellaetal2000,daw2000,wijnandsetal2002_bepposax}. The
source was very dim in quiescence, with a luminosity close to or lower
than $10^{32}$ \Lunit. Due to the limited angular resolution of {\it
BeppoSAX}, doubts were raised as to whether the source detected by
this satellite was truly SAX J1808.4--3658 or an unrelated field
source \citep{wijnandsetal2002_bepposax}. \citet{campanaetal2002}
reported on a quiescent observation of the source performed with {\it
XMM-Newton} which resolved this issue. They detected the source at a
luminosity of $5\times10^{31}$ \Lunit~and found that the field around
SAX J1808.4--3658 is rather crowded with weak sources. Two such
sources are relatively close to SAX J1808.4--3658 and might have
conceivably caused a systematic positional offset during the {\it
BeppoSAX} observations of SAX J1808.4--3658.

After it was found that the source had become active again in January
2000 \citep{vanderklisetal2000}, a Director's Discretionary Time (DDT)
observation request was submitted to the director of {\it XMM-Newton}
to study the outburst X-ray spectrum of SAX J1808.4--3658. This
request was granted and on March 6, 2000, {\it XMM-Newton} observed
the source.  Due to the extreme variability of the source during its
2000 outburst \citep{wijnandsetal2001_rxte}, the {\it XMM-Newton}
observation was performed during times when the source had very low
luminosities (see \S~\ref{section:observation}). Because of this the
DDT observation was considered to be in conflict with an approved
Cycle 1 observation on this source with the purpose of studying the
quiescent X-ray properties of SAX J1808.4--3658 \citep[see][for the
results of that observation]{campanaetal2002}. Consequently, the DDT
observation was not made public until recently.  In this paper I
discuss this DDT observation during which the source was conclusively
detected at a luminosity of $\sim2\times10^{32}$ \Lunit.

\section{Observation and analysis \label{section:observation}}

\begin{figure}[t]
\begin{center}
\begin{tabular}{c}
\psfig{figure=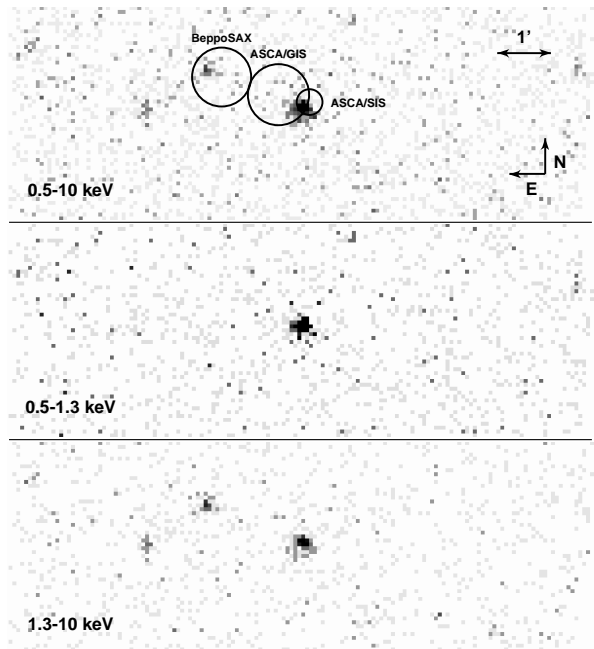,width=8cm}
\end{tabular}
\figcaption{\label{fig:image} {\it XMM-Newton}/MOS1 images of the
region near SAX J1808.4--3658, which is the brightest source in the
images. North is up and East is to the left. The {\it BeppoSAX}, {\it
ASCA}/GIS, and {\it ASCA}/SIS $1\sigma$ error circles of the sources
detected by those instruments are shown
\citep{wijnandsetal2002_bepposax}. The energy band corresponding to
each image is shown on the bottom left.}
\end{center}
\end{figure}

Because a bright X-ray source was expected during the 2000 outburst,
the {\it XMM-Newton}/EPIC-pn camera was used in the timing mode to
eliminate pile-up problems and to study the pulsations. However, due
to the faintness of the source and the high background when using this
mode, the resulting data are not optimal for studying the source
spectrum at the observed low luminosities. Therefore, only the data
obtained with the two metal oxide semiconductor (MOS) cameras will be
discussed. The MOS1 camera was used in the Full Frame mode, but the
MOS2 was in the Small Window mode (again to limit the anticipated
pile-up). For both cameras the medium filter was used.  The source was
not detected in the RGS instruments so that data will not be discussed
further. The log of the observation and the data used in this paper
are listed in Table~\ref{tab:log}.

The data were analysed using the Standard Analysis System (SAS),
version 5.3. The observation was split into two observation
identifiers (IDs; see Tab.~\ref{tab:log}) and it was found that the
MOS1 science data were all present in the first ID (0119940201), but
that the housekeeping data were split over the two IDs (the MOS2
science data {\it and} housekeeping data were both split, in a
consistent manner, over the two IDs). This discrepancy caused the
standard processing of the data to create only about half the
available MOS1 science data. To obtain the full amount of MOS1 science
data, the SAS task {\it emchain} was executed but after the
housekeeping files for the MOS1 were merged\footnote{Using the threads
listed at {\tiny
http://wave.xray.mpe.mpg.de/xmm/cookbook/EPIC\_PN/merge\_odf.html}}
and using the latest calibration files.  Those newly created event
list files (also for the MOS2) were used in the subsequent analysis.
At the end of the observation, a strong background flare occurred. The
data in which the total count rate (all CCDs and no extra filtering)
exceeded 5 and 4 counts per second (using 100 second bins) for the
MOS1 and MOS2 data, respectively, were excluded from the analysis. The
resulting total live time of the central CCD (the CCD on which SAX
J1808.4--3658 is located) is listed in Table~\ref{tab:log}.

The MOS1 image\footnote{The use of the Small Window mode for the MOS2
camera limited the size of the image around SAX J1808.4--3658 and
therefore this image is not displayed.} in the 0.5--10 keV energy band
is shown in Figure~\ref{fig:image} (top) where we clearly see that SAX
J1808.4--3658 was detected. This source is the brightest source on the
central CCD, strongly indicating that during the {\it BeppoSAX} 2000
outburst observations most of the detected flux originated from SAX
J1808.4--365 and not from an unrelated field source as was suggested
by the {\it BeppoSAX} data
\citep{wijnandsetal2002_bepposax}.  \citet{campanaetal2002} reached
similar conclusions using a quiescent observation of the source and
they suggested that the systematic offset of the {\it BeppoSAX}
position with regards to that of SAX J1808.4--3658 might have been
caused by the presence of two faint sources close to SAX J1808.4--3658
\citep[Fig.~\ref{fig:image}, top panel; see also Fig.~1 in
][]{campanaetal2002}. Moreover, due to the source faintness during the
{\it BeppoSAX} observations, only the Medium Energy Concentrator
Spectrometer produced useful data \citep{wijnandsetal2002_bepposax}
and this instrument was only sensitive in the energy range 1.3--10
keV. Therefore, images of the data in the 0.5--1.3 keV and the 1.3--10
keV energy bands were made (Figs.~\ref{fig:image}, middle and bottom
panel) and SAX J1808.4--3658 is detected in both energy ranges. In
contrast, the two extra sources are only detected in the 1.3--10 keV
band, demonstrating that the fractional flux contribution of the two
other sources to the combined flux increases with photon energy. These
differences in source spectra add further evidence to the suggestion
that those two extra sources might have caused the systematic offset
in the {\it BeppoSAX} observations. It should be noted that the {\it
BeppoSAX} fluxes quoted by \citet{wijnandsetal2002_bepposax} and
\citet{stellaetal2000} for SAX J1808.4--3658 are
likely close to its true flux. A small contamination of the fluxes by
the two close-by sources is likely, but those sources are considerably
less luminous than SAX J1808.4--3658 and therefore the contamination
should be small.

\begin{figure}[t]
\begin{center}
\begin{tabular}{c}
\psfig{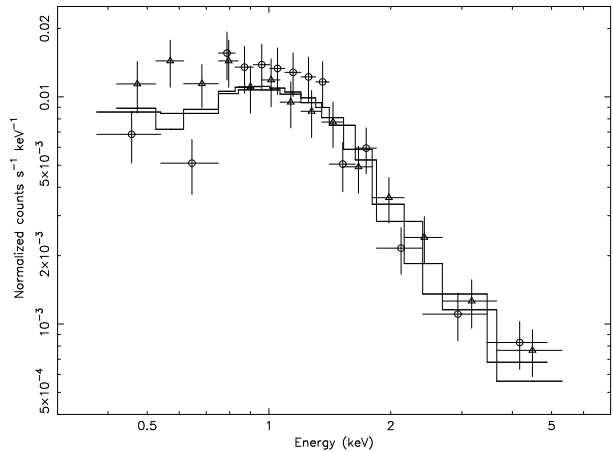}
\end{tabular}
\figcaption{\label{fig:spectrum} {\it XMM-Newton}/MOS1 (circles) and
MOS2 (triangles) spectra of SAX J1808.4--3658 during the 2000 outburst.
The solid lines represent the best power-law fit to the data.}
\end{center}
\end{figure}

The X-ray source spectra were extracted using a circle with a radius
of 20$''$ on the position of SAX J1808.4--3658. For the MOS1 camera,
the background spectrum was extracted using a circle with a radius of
200$''$ on the same position, but excluding the detected point sources
in this region. Because of the use of the Small Window mode, only a
limited field could be use to extract the background spectrum for the
MOS2 camera (i.e., an annulus was used on the source position with an
inner radius of 30$''$ and an outer radius of 50$''$)\footnote{The
accuracy of the background subtraction was checked by using background
regions located on the other, fully read-out, CCDs. The fluxes were
always within 6\% of each other regardless of the background used, and
the spectral parameters within 2\%. To avoid systematic uncertainties
due to the different responses and different offsets of the individual
CCDs, the background obtained from the central CCD was
preferred.}. The RMF and ARF files were created with the SAS tools
{\it rmfgen} and {\it arfgen}. The spectra obtained were then grouped
using the FTOOL {\it grppha} into bins with a minimum of 20 counts per
bin to validate the use of the $\chi^2$ statistics.  The MOS1 and MOS2
spectra are shown in Figure~\ref{fig:spectrum} and were fitted
\citep[using XSPEC version 11.1.0;][]{arnaud1996} simultaneously using
the same model (see Tab.~\ref{tab:spectrum} for the fit
parameters). The column density \nh~ was allowed to float and the
value obtained was always consistent with the value
($1.22\times10^{21}$ cm$^{-2}$) inferred from the $A_{\rm v}$ measured
by \citet{wangetal2001} \citep[and using the relation between \nh~and
$A_{\rm v}$ from][]{ps1995}. The spectrum could be fitted with a
power-law model with index of 2.2\pp0.3 and a 0.5--10 keV luminosity
of $1.7 \times 10^{32}$ \Lunit~ \citep[assuming a distance of 2.5
kpc;][]{intzandetal2001}. A black body model could not fit the data
accurately ($\chi^2$/degrees of freedom = 88.7/25). A neutron star
atmosphere model \citep[that of][]{zps1996} could fit the data but
with a relatively high temperature $kT$ of $\sim$0.2 keV (for an
observer at infinity) and a neutron star radius of 4.5\pp0.1 km (as
measured on the surface and using a neutron star mass of 1.4 \mzon).

The X-ray spectra of other quiescent neutron star transients can
sometimes be described by a two component model \citep[a soft thermal
component below 1 keV and a power-law-like component above a few keV;
e.g.,][]{asaietal1996,asaietal1998} but with the soft component
dominating the spectrum (although the power-law component can
occasionally contribute nearly half the 0.5--10 keV flux of the
source). Although not required by the data, the spectra were also
fitted using such a two component model, with either a black body or
an atmosphere model for the soft component. With these models, the
spectra could be accurately fitted (see Tab.~\ref{tab:spectrum}),
although the atmosphere plus power law combination was very unstable
and the errors on the parameters are therefore relatively large. The
temperatures obtained for the soft component were similar to those
obtained for other quiescent neutron star transients. The flux
contribution of the soft component to the 0.5--10 keV flux was only
$\sim$25\% of the total flux in contrast to the other systems in which
the soft component dominates.

\section{Discussion}

This paper discusses the {\it XMM-Newton} observation of the
accretion-driven millisecond X-ray pulsar SAX J1808.4--3658 performed
during its 2000 outburst. Similar to the {\it BeppoSAX} observations
performed around the same time (one of those observations was on the
same day as the {\it XMM-Newton} one), the {\it XMM-Newton}
observation revealed only a weak X-ray source with a 0.5--10 keV
luminosity of $\sim2\times10^{32}$ \Lunit~at the position of SAX
J1808.4--3658. Despite its weakness, the source was the brightest one
on the central CCD proving that the {\it BeppoSAX} source detected by
\citet{stellaetal2000} and \citet{wijnandsetal2002_bepposax} is indeed
SAX J1808.4--3658 and not an unrelated field source. The systematic
off-set between the measured and the true position of SAX
J1808.4--3658 in those {\it BeppoSAX} observations can likely be
explained by two nearby field sources. These sources might have also
contaminated the flux assigned to SAX J1808.4--3658 \citep[see
also][]{campanaetal2002}.

\citet{campanaetal2002} reported on a quiescent observation of SAX
J1808.4--3658 performed with {\it XMM-Newton} about a year after the
end of the 2000 outburst (the spectral fit parameters reported by
\citet{campanaetal2002} are also listed in Tab.~\ref{tab:spectrum} for
comparison). They found that the source had a luminosity of only
$\sim5 \times 10^{31}$
\Lunit, about a factor of 4 lower than what was measured during the
2000 outburst observation. This demonstrates that at very low
luminosities, SAX J1808.4--3658 can exhibit variability and indicates
that the source probably displays variability in its quiescent state.
However, caution is advised because the 2000 outburst observation was
performed when the source exhibited violent behavior
\citep{wijnandsetal2001_rxte}. Therefore, it is still possible that
in ``true'' quiescence the source will be observed consistently at the
low level reported by \citet{campanaetal2002}.  The spectrum of SAX
J1808.4--3658 during the 2000 outburst appears softer than its
quiescent spectrum: when the spectrum is fitted with a power-law
model, the photon index is 2.2\pp0.3 during the 2000 outburst
vs. 1.5$^{+0.2}_{-0.3}$ in quiescence. The fact that the 2000 outburst
spectrum can be accurately fit with a neutron star atmosphere model or
with a combination of a thermal plus power-law component
\citep[in contrast with the quiescent spectrum;][]{campanaetal2002},
also suggests a difference in the source spectra between the two
epochs. It is interesting to note that in the two-component model, the
photon index during the 2000 outburst observation is very similar to
that measured during the quiescent observation, suggesting that the
shape of the power-law component might not have changed considerably
between the different epochs (although the flux of this component was
still a factor of $\sim$3 higher during the 2000 outburst than in
quiescence).  Due to the limited statistics of the data, the exact
degree of spectral variability cannot be determined nor can the exact
cause of these possible variations be established.

During the 2000 outburst, the source fluctuated in luminosity by over
3 orders of magnitude on timescales of days
\citep{wijnandsetal2001_rxte}.  However, the {\it BeppoSAX}
observations reported by \citet{wijnandsetal2002_bepposax} provided
only rough estimates of the flux of SAX J1808.4--3658 between 2000
March 5 and 8.  With the {\it XMM-Newton} detection of the source on
2000 March 6, we now have a clearer picture of how dim the source
could become during certain phases of its 2000 outburst.  As stated by
\citet{wijnandsetal2001_rxte}, 
the large luminosity variations observed during the 2000 outburst are
difficult to understand as due to similarly dramatic variations in the
mass accretion rate. It is more likely that only modest variations in
the accretion rate can trigger transitions between two significantly
different luminosity states. For example, centrifugal inhibition of
accretion by the neutron star's magnetic field is expected below a
certain critical accretion rate \citep[the 'propeller
regime';][]{is1975} and small but erratic variations in the accretion
rate around this critical rate could in principle give rise to the
enormous luminosity swings observed during the 2000 outburst.  When
the source is in this propeller regime, accretion is inhibited, but it
is evident from the large brightness fluctuations during the 2000
outburst, that a considerable amount of matter was still available in
the accretion disk. According to \citet{campanaetal2002}, a pure
propeller contribution is ruled out in quiescence since this mechanism
is expected to stop operating at luminosities below $10^{33}$
\Lunit~because the source should turn on then as a radio
pulsar. \citet{stellaetal2000} and \citet{campanaetal2002} suggested
that a possible explanation for the quiescent flux is the emission
from the shock front between the relativistic wind of the radio pulsar
and the matter out-flowing from the companion star. Therefore, the
flux during the 2000 outburst observation could be higher than that
observed in quiescence because of the large amount of matter still
present close to the neutron star. 

\citet{db2003} suggested an
alternative explanation for the quiescent emission of SAX
J1808.4--3658 in which this emission is produced by direct dipole
radiation from the radio pulsar. The quiescent X-ray luminosity and
spectral shape of SAX J1808.4--3658 are indeed consistent with those
observed from several field millisecond radio pulsars \citep[see][for
a recent review]{bp2002}, but \citet{grindlay2002} found that the
millisecond pulsars in the globular cluster 47 Tuc have a
predominantly soft spectral shape suggestive of a thermal origin in
contrast to what is observed for SAX J1808.4--3658. Conclusive proof
for an active radio pulsar in SAX J1808.4--3658 would come from the
detection of radio pulsations during quiescence, although such a
detection might be inhibited by the ambient matter still present in
the system \citep[see also the discussion in][]{db2003}.

\citet{campanaetal2002} 
noted that the quiescent properties of SAX J1808.4--3658 are
remarkably different from those observed for other quiescent neutron
star systems: (a) its 0.5--10 keV luminosity is the lowest observed so
far for any neutron star system, and (b) its spectrum is dominated by
a power-law component instead of a thermal component.  The quiescent
emission of neutron star X-ray transients is most often explained by
thermal emission from the neutron star surface releasing the heat
deposited in the crust and core of the neutron star during outburst
\citep[see][ and references therein]{bbr1998}.
In this model, the exact luminosities of the systems should depend on
their time-averaged accretion rates
\citep{campanaetal1998,bbr1998}.
Due to the low peak luminosity of SAX J1808.4--3658, \citet{bbr1998}
predicted that this source should be rather faint in quiescence. At
first sight the low detected quiescent luminosity is consistent with
this prediction, but no strong evidence could be found for a thermal
component in the quiescent spectrum obtained. This indicates that the
thermal luminosity of this source is very low, implying a rapidly
cooling neutron star which would require enhanced core cooling
processes to explain \citep{campanaetal2002}.

The combination of the low time-averaged accretion rate and the
possibility of rapid core cooling might also be able to explain why
the quiescent spectra of SAX J1808.4--3658 is dominated by the
power-law component.  If in the 'ordinary' quiescent systems the
thermal component were to drop to low luminosity levels similar to
those of SAX J1808.4--3658, then their quiescent spectra would be
dominated by the power-law component (assuming that this component
remains unchanged), just like in SAX J1808.4--3658. However, it
remains to be determined if the power-law components in the different
types of systems are due to the same mechanism, or if the quiescent
spectrum of SAX J1808.4--3658 is due to some other process, likely
related to the different configuration and/or magnetic field strength
of the neutron star in SAX J1808.4--3658.  To get more insight into
the nature of the unique quiescent properties of SAX J1808.4--3658,
more neutron star transients must be detected in
quiescence. Particularly interesting systems are the two other, very
recently discovered, accretion driven millisecond X-ray pulsars XTE
J1751--305
\citep{markwardtswank2002} and XTE J0929--314
\citep{rss2002}. It would be interesting to compare their quiescent
properties with those of SAX J1808.4--3658 and correlate possible
similarities and/or differences in quiescence with the outburst
properties of those sources.

\acknowledgments

This work was supported by NASA through Chandra Postdoctoral
Fellowship grant number PF9-10010 awarded by CXC, which is operated by
SAO for NASA under contract NAS8-39073.

\begin{deluxetable}{cccc}
\tablecolumns{4}
\tablewidth{0pt}
\tablecaption{Log of the {\it XMM-Newton} observation \label{tab:log}}
\tablehead{
Observation ID      & Time of observation            & Instrumental modes$^a$ & Live time$^b$\\
                    & (UTC, 6 March 2000)            &                        & (ksec)   }
\startdata
0119940201          &            18:29 - 23:53       & MOS1 FF/Medium         & 16.1 \\
                    &            17:09 - 20:24       & MOS2 SW/Medium         & 11.4 \\
0119940501          &            21:53 - 23:51       & MOS2 SW/Medium         &  3.9 \\
\enddata

\tablenotetext{a}{FF is Full Frame mode and SW is Small Window
mode. Medium indicates that the medium filter was used during the
observations.}

\tablenotetext{b}{Live time of the CCD on which SAX J1808.4--3658 is
located, after elimination of the background flares.}

\end{deluxetable}

\begin{deluxetable}{rcccc}
\tablecolumns{5}
\tablewidth{0pt}
\tablecaption{Spectral parameters$^a$ \label{tab:spectrum}}
\tablehead{
Model               & \nh                    & Index/$kT$                 & Flux               & $\chi^2_{\rm red}$  \\
                    & ($10^{21}$ cm$^{-2}$)  &                            & ($10^{-13}$ \funit)&                     }
\startdata										                         
\multicolumn{5}{c}{Single component model}\\						                         
\hline											                         
Power law: 2000     & $1.2^{+0.5}_{-0.7}$    & 2.2\pp0.3                  & 2.3                & 1.3                      \\
           2001     & $0.3^{+0.7}$           & $1.5^{+0.2}_{-0.3}$        & 0.62               & 1.1                           \\ 
Black body: 2000    & $<0.1$                 & $0.37^{+0.05}_{-0.04}$ keV & 1.1                & 3.5                       \\
            2001    & $0.0^{+0.3}$           & $0.52^{+0.11}_{-0.08}$ keV & 0.29               & 3.2                           \\
Atmosphere$^b$: 2000& $<1$                   & $0.20^{+0.03}_{-0.05}$ keV & 3.4                & 1.3                       \\
                2001& $0.4^{+1.1}$           & $0.27^{+0.0}_{-0.02}$ keV  & 0.24               & 4.4                         \\
\hline									                             	                 
\multicolumn{5}{c}{Multi component model, only for the 2000 outburst observation}\\
\hline									                             	                 
Black body + power law$^c$ & 1.3\pp0.5       & 0.2\pp0.1 keV              & 2.6                & 1.2                     \\
                       &                     & 1.6\pp0.3                  &                    &                              \\
Atmosphere + power law$^d$ & 1\pp1           & $0.10^{+0.11}_{-0.06}$ keV & 2.6                & 1.2                     \\
                       &                     & $1.4^{+0.7}_{-0.4}$        &                    &                              \\ 
\enddata

\tablenotetext{a}{The errors are for 90\% confidence levels. The
fluxes are unabsorbed and in the 0.5--10 keV energy range. For the
single component models, the fit parameters are given for the 2000
outburst and also for the 2001 quiescent observations as reported by
\cite{campanaetal2002}. For the two-component models, only the 2000
outburst results are given. }

\tablenotetext{b}{The atmosphere model 
by \citet{zps1996} was used for the 2000 outburst observations and
with a distance of 2.5 kpc and a neutron star mass of 1.4
\mzon. \citet{campanaetal2002} used the hydrogen atmosphere model by
\cite{gansickeetal2002} to obtain the fit result for the 2001
quiescent observation.}

\tablenotetext{c}{The black body flux was $0.6$ and the power law flux
$2.0\times 10^{-13}$ \funit~(0.5--10 keV).}

\tablenotetext{c}{The flux in the atmosphere component was $0.7$ and
the power law flux $1.9\times 10^{-13}$ \funit~(0.5--10 keV).}

\end{deluxetable}

\end{document}